# Gaseous Scissor-mediated Electrochemical Exfoliation of Halogenated MXenes and its Boosting in Wear-Resisting Tribovoltaic Devices

Qi Fan[1,2,#], Minghua Chen[1,#], Longyi Li[2,3,#], Minghui Li[4,5], Chuanxiao Xiao[4,5], Tianci Zhao[6], Long Pan[7], Ningning Liang[6], Qing Huang[1,8], Laipan Zhu[2,3]*, Michael Naguib[9], Kun Liang[1,2,8]*

[1]Zhejiang Key Laboratory of Data-Driven High-Safety Energy Materials and Applications, Ningbo Key Laboratory of Special Energy Materials and Chemistry, Ningbo Institute of Materials Technology and Engineering, Chinese Academy of Sciences, Ningbo 315201, P.R. China

[2]University of Chinese Academy of Sciences, 19 A Yuquan Rd, Shijingshan District, Beijing 100049, P. R. China

[3]Beijing Key Laboratory of Micro-nano Energy and Sensor, Center for High-Entropy Energy and Systems, Beijing Institute of Nanoenergy and Nanosystems, Chinese Academy of Sciences, Beijing 101400, P. R. China

[4]Ningbo Institute of Materials Technology and Engineering, Chinese Academy of Sciences, Ningbo 315201, P. R. China

[5]Ningbo New Materials Testing and Evaluation Center Co., Ltd, Ningbo 315201, P. R. China

[6]School of Physics and Optoelectronic Engineering, Beijing University of Technology, Beijing 100124, P. R. China

[7]Key Laboratory of Advanced Metallic Materials of Jiangsu Province, School of Materials Science and Engineering, Southeast University, Nanjing 211189, P. R. China

[8]Qianwan Institute of CNITECH, Ningbo 315201, P.R. China

[9]Department of Physics and Engineering Physics, Tulane University, New Orleans, LA 70118, USA

E-mail: zhulaipan@binn.cas.cn; kliang@nimte.ac.cn

**Abstract:**

Two-dimensional transition metal carbides (MXenes), especially their few-layered nanosheets, have triggered burgeoning research attentions owing to their superiorities including extraordinary conductivity, accessible active surface, and adjustable processability. Molten salts etching route further achieves their controllable surface chemistry. However, the method encounters challenges in achieving few-layer structures due to more complex delamination behaviors. Herein, we present an efficient strategy to fabricate Cl- or Br-terminated MXene nanoflakes with few-layers, achieved by electrochemical intercalation of Li ions and concomitant solvent molecules in the electrolyte solution, with gaseous scissors (propylene molecules) to break up interlayer forces. By controlling cut-off voltages, the optimal protocol results in nanosheets with an ultrahigh yield (~93%) and preserved surface chemistry. The resultant MXenes dispersions were employed as lubricants to enhance tribovoltaic nanogenerators, where $Ti_3C_2Br_2$ displayed superior electrical output. These findings facilitate the understanding of MXenes' intrinsic physical properties and enable the nanoengineering of advanced electronic devices.

## Introduction

Since their initial report in 2011, transition metal carbides or nitrides, known as MXenes, particularly their mono- or few-layered nanosheets, have distinguished themselves among numerous two-dimensional (2D) materials owing to their unique properties[1,2]. Amalgamating the superiorities of both adjustable structural components and versatile physicochemical properties, MXenes can be synthesized using a programmable structural editing protocol to achieve extraordinary performance in a wide range of customized applications[3,4]. Surface chemistry modification plays a crucial role in tailoring the band structures and work functions of MXenes to achieve a broad range of tunable properties, especially in electronics[5] and optoelectronics[6-9]. The pristine surface groups of MXene nanosheets are primarily determined by the etching process. The conventional wet chemistry etching route selectively removes the A layers from MAX phases using hazardous fluoride ion-containing acids, resulting in a mixture of -OH, -O and -F terminations on the surface of MXene nanosheets[10]. Although these electronegative terminations impart the nanomaterials with favorable aqueous processability without the need for surfactants or binders, further development of MXenes-based functional materials is challenged by their mediocre and uncontrollable adjustment. This limitation persisted until the introduction of the molten salts etching route, which allowed for full control of the surface terminations[11-13]. Through substitution and elimination reactions in molten salts, a variety of affluent terminations such as -Cl, -Br, -I, -S, -Se, and -Te, can be successfully synthesized, creating new design opportunities for the structural and electronic properties of MXenes, termed as MS-MXenes[3,9,14]. However, the higher adhesion energy values or stronger surface hydrophobicity of the terminations hinder the separation of layers with covalent surface modifications, thereby limiting their application in micro-nano electronics such as tribovoltaic nanogenerators (TVNGs)[9,15,16].

The intercalation behavior of external species is crucial throughout the production of MXene nanomaterials, encompassing stages from etching and exfoliation to delamination and post-treatment[4,17]. Ideally, numerous intercalants, including cations, solvent molecules, and organic molecules, can intercalate into and expand the subnanometer 2D galleries via weakening interflakes interactions, potentially delaminating multi-layer MXenes into mono-layer or few-layer nanoflakes[18-20]. However, current exfoliation strategies face several challenges, such as reliance on hazardous reagents (e.g., n-butyllithium and sodium hydride), and low yields for high-quality single-layer flakes[9,13]. Other methods, like the introduction of tetramethylammonium cations, result in brittle and rigid few-layer membranes due to improper intercalation degree[15]. Although the LiF-involved molten salt etching route facilitates the delamination process by enhancing the interaction between -F groups and tetrabutylammonium hydroxide (TBAOH) intercalants, it runs counter to fluoride-free intention and limits the exfoliation yield to approximately 20%[16]. Recently, Gogotsi et al. reported a more efficient strategy using LiCl as a delaminating agent and N-methylformamide (NMF) as an optimized solvent for further swelling of powders[21]. However, scaling up remains challenging due to the complexity of the process, the difficulty in removing organic solvents, and the lack of universality or efficiency for other categories of MS-MXenes.

In response to these issues, electrochemical exfoliation has emerged as one of the most promising and convenient strategies to prepare MS-MXenes nanosheets with high quality and high yield, suitable for targeted applications[22-24]. This powerful synthetic toolkit typically involves the electrochemical intercalation of solvated cations or anions in the electrolyte, followed by a sonication and exfoliation process[25]. Combined with easy procedures and mild reaction conditions, the method offers a higher degree of control over the intercalation process by regulating cut-off voltage and discharge current[26]. Intriguingly, solvent molecules in electrolyte often co-intercalate into layered materials, significantly expanding the interlayer space. For instance, $Li(PC)_n^+$ solvated ions are responsible for the intensive exfoliation of graphite layers, due to the formation of an ineffective solid electrolyte interface film and the continuous decomposition of solvent molecules within the interlayer space[27-29]. It has been reported that the electrolyte solvent profoundly affects the pseudocapacitive charge storage of $Ti_3C_2$-MXene, with carbonate solvent (e.g., PC) contributing to a maximized capacitance[30].

In this work, we present a powerful and versatile structural-editing protocol for gas molecular scissor-mediated electrochemical exfoliation of MXenes synthesized via molten salts etching route, based on solvation-co-intercalation-decomposition of Li(PC)$_n^+$ ions. By regulating the cut-off voltage during the synthesis process, we optimized the yield of few-layer MXene nanosheets to an impressive 93%, while retaining the pristine surface chemistry, such as -Cl or -Br terminations. In-situ X-ray diffraction (XRD), in-situ differential electrochemical mass spectrometry (DEMS), and ex-situ Fourier transform infrared spectroscopy (FT-IR) measurements intuitively corroborated the crucial appearance and role of gaseous scissor to break up the intense interactions between layers of MS-MXenes. Subsequently, MXenes with various terminations (-F, -Cl, or -Br) in water or ethanol dispersions were employed as lubricates to improve the output current density of TVNGs, with optimal performance founded in $Ti_3C_2Br_2$, thanks to an effective contact electrification process.

**Results and Discussion**

**Materials Characterizations**

Fig. 1 illustrates the gas molecular scissor-mediated electrochemical intercalation-based exfoliation process for $Ti_3C_2Cl_2$ or $Ti_3C_2Br_2$ MXene nanosheets and their optimal applications in semiconductor-semiconductor tribovoltaic nanogenerator (SS-TVNG). Initially, multi-layered $Ti_3C_2Cl_2$ or $Ti_3C_2Br_2$ were successfully synthesized by selectively removing Al atoms in MAX phase by Lewis acid molten salts etching route. The representative accordion-like structures and their energy dispersive spectroscopy (EDS) images (Supplementary Fig. 1a-f) demonstrate the uniform distribution of Cl or Br elements as functional terminations across all MXene lamellae. XRD patterns (Supplementary Fig. 2) further verify the high-quality MXene prepared by molten salts and the (002) interlayer spacing, which is 1.10 nm for $Ti_3C_2Cl_2$ and 1.20 nm for $Ti_3C_2Br_2$. Next, the electrochemical intercalation-assisted exfoliation route of MS-MXenes were performed in a half-cell configuration, where MS-MXenes pellets served as working electrodes and Li metal acted as counter electrode (Supplementary Fig. 3a). Initially, a commercial electrolyte, 1 M LiPF$_6$ in EC: DEC: EMC=1:1:1(Vol%), was used to conduct the lithiation process of MS-MXenes. However, there was no shift of (002) peak to lower angles, and numerous bulk materials with by-product on their surfaces were observed, indicating ineffectiveness of this electrolyte and its correlation with the electrochemical intercalation behavior of Li ions (Supplementary Fig. 4a, b). Consequently, various solvents were screened to decipher the electrochemical intercalation process of solvated ions and their significant roles in assisting exfoliation of 2D nanosheets, including lithium bis(trifluoromethylsulfonyl)amine (LiTFSI) in propylene carbonate (PC), dimethyl sulfoxide (DMSO), 1,2-dimethoxyethane (DME), dimethyl carbonate (DMC), acetonitrile (ACN), and 2-methyltetrahydrofuran (2Me-THF).

The prepared configurations were connected to a potentiostat for effectively controlling galvanostatic discharge currents and tuning cut-off voltages. After sonication to facilitate exfoliation, the products were centrifuged at 5000 rpm to eliminate the sediment. The successful preparation of few-layer nanosheets can be thoroughly characterized as follows (Fig. 2a-m). Digital photographs of the collected delaminated MS-MXene solutions disclosed their colloidal stability, with the suspension of PC-derived MS-MXene remaining stable even after 4 weeks, in stark contrast to the precipitation of others within 2 hours (Supplementary Fig. 5a). As shown in Supplementary Fig. 5b, only the PC-based electrolyte resulted in the typical (002) peak of few-layer MXene paper among numerous exfoliated products, indicating that PC serves as a desirable alternative for intercalation and swelling of MS-MXene. To further elucidate the evolution of van der Waals (vdW) gaps in a PC-based electrolyte, XRD patterns of multi-layered $Ti_3C_2Cl_2$ or $Ti_3C_2Br_2$, intercalated compounds, and delaminated papers were recorded, showing dramatically enlarged interlayer spacing and weakened interlayer vdW interactions (Fig. 2e and Supplementary Fig. 6). The process is visually demonstrated in photographs where, compared with the original pellet, heavy swelling was observed with significant volume expansion (~2-3 times) for the intercalated compounds. The prepared few-layer $Ti_3C_2Cl_2/Ti_3C_2Br_2$ nanosheets exhibit an ultrathin 2D flake morphology of MXene in the transmission electron microscopy (TEM), atomic force microscopy (AFM) images, and scanning electron microscopy (SEM), with lateral sizes around hundreds of nanometers (Fig. 2a-d and f). The expected hexagonal crystalline symmetry is visible in the selected area electron diffraction (SAED) patterns (Fig. 2a, b), highlighting the effectiveness of this mild protocol in preserving the highest quality of the pristine MS-MXenes. Additionally, AFM images (Fig. 2c,

d) further manifested their ultrathin structure, with average thicknesses of ~2.15 nm for $Ti_3C_2Cl_2$ nanosheets and ~2.32 nm for $Ti_3C_2Br_2$ nanosheets, which indicates obtained flakes consist of only two-layers. Thanks to the mild and effective elimination process, a free-standing and flexible paper of few-layer $Ti_3C_2Cl_2/Ti_3C_2Br_2$ can be obtained (Fig. 2j). More importantly, the maintained surface chemistry, with evenly distributed Cl or Br elements in cross-sectional mapping images, makes them promising candidates for uncovering the intrinsic physiochemical properties of halogen-terminated MXene flakes (Fig. 2g-i and Supplementary Fig. 8a-c). Unlike some methods where harsh conditions or physical mechanical exfoliation can lead to the loss or alteration of desirable surface terminations (like Cl or Br), this electrochemical approach preserves these functional groups. This retention is crucial as it maintains the material's surface chemistry, enhancing its properties for specific applications.

In brief, the utilization of PC solvent plays a critical role in delaminating MS-MXenes nanosheets with an excellent yield of 93 %. This efficiency translates to higher yields and improved quality of the nanoflakes, with fewer defects and impurities. Numerous coin cells can operate on the battery testing system in the meantime to guarantee satisfying product output (Fig. 2k). Further scaling up preparation process can be achieved in a two-electrode Swagelok cell by handling 3.5 g of multilayer MS-MXene powders, which is also suitable for these laboratories lack of battery assembly facilities (Fig. 2l and Supplementary Fig. 3b, c). A digital photograph of the collected delaminated MXenes solution is shown in Fig. 2m, exhibiting its superior yield. Theoretically, the scale of preparation can be further expanded with larger cells and equipment supports. The ability to produce MXenes at high yield not only reduces the cost per unit but also accelerates the research and development cycle for new applications, giving industries confidence to invest in MXene-based technologies. This scalability is crucial for transitioning from laboratory research to commercial and industrial utilization.

**Mechanism of Electrochemical Intercalation-based Exfoliation Strategy**

The crucial role of the PC-based electrolyte in electrochemical intercalation-assisted exfoliation of MS-MXenes could be elucidated through in-situ XRD, in-situ DEMS, and ex-situ FT-IR measurements. We conducted in-situ XRD measurements on the multi-layers $Ti_3C_2Cl_2$ electrode to monitor the successive variation of interlayer spacing induced by intercalated solvated Li$^+$ (here, Li(PC)$_n^+$) as the voltage decreased to specific levels (Fig. 3a, b). Three distinct stages were identified, intercepted by voltages of 0.86, 0.5, and finally 0.01 V (Fig. 3a). As shown in the XRD patterns (Fig. 3b), Stage 1 is characterized by a decreasing low-angle peak area as the discharge voltage drops from 3.0 to 0.86 V, which can be ascribed to the intercalation of abundant Li$^+$ ions. Strikingly, Stage 2 exhibits a smooth plateau between 0.86 V and 0.5 V, during which the XRD peaks gradually fade, implying a temporarily reduced degree of order in the intercalated compound or the emergence of an amorphous phase from decomposed electrolyte. In stage 3, the patterns remain consistent, suggesting the unencumbered migration of Li$^+$ in the overlarge interlayer space. It is noteworthy that, at 0.01 V, the (002) peak reappears after cleaning up byproducts of electrolyte decomposition, corresponding to the remarkably left-shifted characteristic peak of Ti$_3$C$_2$Cl$_2$ (Fig. 2e). To delve deeper into the electrochemical reactions, DEMS was utilized to detect the products formed during the charge-discharge process. Remarkably, propylene molecules were detected at around 0.86 V, in accordance with the process of gradually disappeared (002) XRD peak at Stage 2 (Fig. 3c). There have been several reports about the electrochemical reduction of PC electrolyte with Li salts[27,29,31], and the decomposition mechanism at particular potential is exhibited in Supplementary Fig. 9. During the electrochemical process, the electrolyte undergoes decomposition, producing gas that plays a critical role in the exfoliation of MXene layers. This gas generation introduces localized pressure between the layers, aiding in their separation without the need for harsh mechanical or chemical treatments. In addition, the decomposed gas effectively infiltrates between the MXene layers, creating an expansion force that promotes swelling. This controlled swelling effect is crucial for achieving a uniform delamination, thus resulting in high-quality few-layer MXenes. Additionally, trace amounts of hydrogen gas from the reduction of residual water in electrode or electrolyte were detected at the beginning of measurement (Supplementary Fig. 10). Furthermore, ex-situ FT-IR characterization was employed to investigate electrolyte decomposition products at different discharge stages (Fig. 3d). The appearance of characteristic peaks for stretching vibration of Li$_2$CO$_3$ at 0.5 V and 0.01

V in the intercalated compounds, forcefully supports the decomposition of the PC-based electrolyte, consistent with known behavior in Li metal battery (Supplementary Fig. 9) [27,29,31]. The mechanism using gas from electrolyte decomposition is inherently scalable. The generation of gas can be controlled by adjusting the electrochemical parameters, providing a flexible method that can be scaled from small lab setups to larger production environments. This method ensures reproducibility across batches, as the uniform gas generation leads to consistent exfoliation results, which is a significant advantage over methods that rely on less controllable mechanical forces.

**Optimization of Cut-off Voltages for High Yield**

Based on the above experimental verifications, an "intercalation-swelling-delamination" mechanism is proposed for producing few-layer $Ti_3C_2Cl_2$ or $Ti_3C_2Br_2$ nanosheets, incorporating with theoretical decomposition reaction (Fig. 4a-c and Supplementary Fig. 9). More intuitively, ex-situ SEM images at specific discharge potential reveal the evolution of the layered structure from compact to aerated, benefiting from the collapse of distinctive interaction between MS-MXenes layers, stimulated by the gaseous propylene "physical scissor". From the systematic research outlined above, it is evident that the selection of cut-off voltage is crucial in determining the exfoliation yield of MS-MXenes by effectively modulating the amount or status of intercalated $Li(PC)_n^+$ species. Exfoliation yields at various cut-off potentials were calculated, revealing that the optimized yield could reach 93 %, further emphasizing the critical role of gaseous molecules generation at 0.5-0.86 V (Fig. 4d). The highly efficient exfoliation process is also highlighted in the video (Supplementary Video 1), demonstrating how the intercalated compounds generate the desired amounts of exfoliated sheets within minutes when immersed in deionized water, without the need for sonication. Noteworthily, the higher exfoliation yield is irrelevant to discharging to lower potential than 0.5 V, where the highly negative zeta potential of -40.4 mV enhances strong electrostatic repulsion between few-layer $Ti_3C_2Cl_2$/$Ti_3C_2Br_2$ nanosheets, ensuring uniform sizes and stable colloidal dispersion for modulating electronic properties (Supplementary Fig. 11a-d). It has been reported that the functional groups on MXene surfaces significantly impact their electronic structure and optical absorption[32,33]. Directly associated with electron charge transfer or electronegativity, the work function (W) can be calculated following the formula of $W_F=h\nu - (E_{cut-off}-E_F)$ from the ultraviolet photoelectron spectroscopy (UPS) characterization results, in sequence of $Ti_3C_2Br_2$ (~3.74 eV), $Ti_3C_2Cl_2$ (~4.06 eV) and $Ti_3C_2T_x$ (~4.49 eV) (Fig. 4e and Supplementary Fig. 12). Furthermore, electrical conductivity of the three types of MXenes films was measured using a four-point probe set-up. The results indicate that $Ti_3C_2Br_2$ ($2.09\times10^3$ S/m) and $Ti_3C_2Cl_2$ ($2.24\times10^4$ S/m) offer more moderate values compared to those obtained from a wetting chemical route ($1.70\times10^5$ S/m) (Fig. 4e). The UV-vis spectra (Supplementary Fig. 13) of delaminated $Ti_3C_2T_x$, $Ti_3C_2Cl_2$, and $Ti_3C_2Br_2$ solutions exhibit evident absorption peaks in the near-infrared (NIR) region, with $Ti_3C_2Cl_2$ and $Ti_3C_2Br_2$ showing a 60 nm red shift to 860 nm, in stark contrast to conventional $Ti_3C_2T_x$. In addition, the good hydrophilicity of conventional $Ti_3C_2T_x$ is confronted with severe oxidation or decomposition issues caused by $H_2O$. Fortunately, modulating surface terminations allows MXene materials to achieve adjustable surface wettability, freeing them from the trade-off between processability and durability in practical applications. Contact angle tests (Fig. 4f) provide clear evidence of the increased hydrophobicity of MXenes with surface groups transitioning from -F, -O, and -OH (~37°) to -Cl (~68°) or -Br (~83°), which is expected to retain the intrinsic properties of multilayer structures and bodes well for applications related to interface interaction. Ethanol contact angles present similar trends with little differences (Fig. 4f).

By focusing on the innovative use of gas from electrolyte decomposition, this mechanism not only improves the efficiency and yield of MXene production but also preserves the structural and chemical integrity necessary for advanced applications. This represents a significant improvement over existing exfoliation strategies, highlighting the potential for more sustainable and economically viable MXene synthesis.

**Effects of different MXenes Lubricants on Output Performance of TVNGs**

When two semiconductors slide past each other, chemical bonds at the interface continuously break and reform. During this interfacial bonding process, energy quanta known as bindingtons are released. If these bindingtons possess sufficient energy, they can excite electron-hole pairs.

These newly generated charge carriers are then separated by the built-in electric field, producing a current that flows between the two ends of the device, connected through an external circuit. The overall process was called as the tribovoltaic effect[34-36]. Contact electrification is the principal mechanism to form TVNGs, which is drastically influenced by the interface interaction and electronic nature of the two contacted semiconductor materials[37,38]. Adjusting the sliding surface carriers and interface wettability provides an effective route to increase electrical output and mitigate wear issues of the TVNGs, by introducing lubricants between the sliding semiconductors. Although $Ti_3C_2T_x$ MXene aqueous dispersion has been utilized to lubricate the sliding friction interface, previous studies often stopped at investigating the inherent surface properties and electronic nature of MXene nanoflakes, defined by the terminations on MXenes surface[39]. Herein, we explored p/n-GaN SS-TVNG applications (Fig. 5a) with different functional groups (including -F, -Cl, and -Br) based MXenes lubricants by recording their electrical output and stability tests. The SS-TVNG operating parameters and MXene dispersions used in experiments are provides in Supplementary Table S1 and Table S2. Electrical output results were provided to assess the enhancement effect of various MXene dispersions at varying concentrations on SS-TVNG output. As shown in Fig. 5b, c and Supplementary Fig. 14, the output of SS-TVNG at dry condition is only about 44 nA, which is significantly increased to microampere level when using MXene lubricants in water or ethanol solutions. It is noteworthy that the output shows a typical dependence on the concentration, peaking at an optimum MXenes concentration. Specifically, at low concentrations (≤1 mg/ml) in both dispersions, the enhancement effects of -Cl or -Br terminations surpass those of -F terminated MXene, with aqueous dispersion providing the best overall performance (Supplementary Fig. 15). To better understand the mechanism of interface lubricant from various MXenes lubricants, the I-V characteristics of the p/n-GaN static heterojunctions with or without MXene droplets were tested, with raw data in Supplementary Fig. 16 and 17. Static interface resistances can be calculated from the current value at 4 V forward bias, exhibiting a remarkably facilitated interface transmission after introducing MXene lubricants, contrasting with 279.80 MΩ in the dry state (Fig. 5d, e). In water solutions, the interface resistances trend follows -Br > -Cl > -F, with higher concentration solution forming a viscous film on sliding interface, reducing the contact area and thus electrical output. Nevertheless, the resistance difference is minor for the ethanol-based lubricants.

All these impressive results demonstrate that $Ti_3C_2Cl_2$ and $Ti_3C_2Br_2$ lubricants reveal a superior enhancement effect on TVNG output compared to $Ti_3C_2T_x$, which can be ascribed by solid-liquid contact friction and electrification between MXene droplets and semiconductor wafers, as depicted in Fig. 5f. Negative charges generated by contact electrification on the semiconductor surface separate from the liquid droplets during sliding, generating current signals in subsequent hard solid-solid p-n contacts, amplified by larger contact angles of $Ti_3C_2Br_2$ and secondly $Ti_3C_2Cl_2$ (Fig. 4f). Based on experimental results, this interface effect dominates in the low concentration range. Figure 5g further illustrates the energy band diagram of the p-n junction in the SS-TVNG to better understand the principle of output enhancement from the perspective of the electric field. The charge separated by discontinuous solid-liquid interfaces forms a triboelectric field in the same direction as the p-n junction, providing more electron-hole pairs than under dry conditions, thereby boosting the total electric field and signal output during SS-TVNG contact friction. Simultaneously, the addition of MXene droplets affects interface resistance, acting as a variable resistor in the equivalent circuit and primarily influencing TVNG output in the high concentration range. Moreover, MXene lubricants also play a crucial role in protecting semiconductor surface and extending the device lifespan, proved by wear resistance tests conducted on dry, dispersed, and MXene dispersion-added SS-TVNGs for over 10000 cycles. Optical images (Fig. 5h and Supplementary Fig. 18) show that interfaces with MXene dispersions exhibit fewer scratches than those with only DI water or under dry conditions.

## Conclusion

In this work, we pioneered an efficient electrochemical exfoliation approach to achieve high-yield for few-layered MXenes terminated with -Cl/-Br groups, by leveraging gaseous propylene molecules as a physical scissor. The results underscore the significance of electrochemistry-modulated "intercalation-swelling-delamination" mechanism, supported by the strategic use of decomposed gas from electrolytes, this approach achieves high yield and maintains crucial surface functionalities. Motivated by the substantial impact of surface groups modification on tailoring electronical and surface properties of MXenes, the obtained $Ti_3C_2Br_2$ dispersion as interface lubricant contribute to high output performance TVNG. The scalability and adaptability of this method open new pathways for industrial applications, setting a new benchmark in material science and reflecting the potential of MXenes to drive technological innovations that meet global demands for sustainable and high-performance solutions.

## Methods

### Materials

$Ti_3AlC_2$ (99%, 300 mesh) were purchased from Jilin 11 Technology Co., Ltd, China. $CdCl_2$ (anhydrous, 99.99%), $CdBr_2$ (anhydrous, 99.99%), KCl (anhydrous, 99.99%), NaCl (anhydrous, 99.99%), lithium bis(trifluoromethylsulfonyl)amine (LiTFSI, 99.99%), propylene carbonate (PC, anhydrous, 99.99%), dimethyl sulfoxide (DMSO, anhydrous, 99.99%), 1,2-dimethoxyethane (DME, anhydrous, 99.99%), dimethyl carbonate (DMC, anhydrous, 99.99%), acetonitrile (ACN, anhydrous, 99.99%) and 2-methyltetrahydrofuran (2Me-THF, anhydrous, 99.99%), HF (49 wt%), tetrabutylammonium hydroxide (TBAOH, ~40 wt% in $H_2O$) were purchased from Aladdin. HCl (36-38%) were purchased from Sinopharm Chemical Reagent Co., Ltd. The p-type and n-type GaN sapphire-based epitaxial wafers were obtained from Jiangsu Wuxi Jingdian Semiconductor Materials Co., Ltd.

### Synthesis of halogenated multi-layers MXenes

$Ti_3AlC_2$ (0.3 g) powder was mixed with $CdCl_2$ in NaCl/KCl salts at the ratio of 1:3:6:6 by using a mortar and pestle. The resulting mixture was heated in an alumina crucible under Ar at 650 °C for 5 h with a ramping rate of 5 °C/min. And $Ti_3AlC_2$ (0.1 g) powder was mixed with $CdBr_2$ at the ratio of 1:8 by using a mortar and pestle. The resultant mixture was heated in an alumina crucible at 650°C for at least 12 h. After washing with HCl or HBr and following DI water, the product was filtrated and dried under vacuum at 110 °C for 3 h to obtain $Ti_3C_2Cl_2$ or $Ti_3C_2Br_2$ powder for electrode materials.

### Synthesis of few-layer halogenated MXenes nanosheets

Few-layer $Ti_3C_2Cl_2$ or $Ti_3C_2Br_2$ nanosheets were synthesized by electrochemical intercalation assisted exfoliation route. At first, a series of electrolyte solutions were prepared, i.e. 1 M LiTFSI in propylene carbonate (PC), dimethyl sulfoxide (DMSO), 1,2-dimethoxyethane (DME), dimethyl carbonate (DMC), acetonitrile (ACN) and 2-methyltetrahydrofuran (2Me-THF). Ca. 0.1 g of synthesized $Ti_3C_2Cl_2$ or $Ti_3C_2Br_2$ powders were compressed into a cylindrical pellet (Φ 14 mm) under 1 MPa. Then, in an Ar-filled glovebox with $H_2O$ and $O_2$ < 0.01 ppm, a coin cell or a two-electrode Swagelok (TMAX-2E) cell was configured with the pellet as work electrode, Li foil (Φ 14 mm) as counter electrode and pp film as a separator in aforementioned electrolyte solutions. The cell was connected to galvanostatic charge-discharge analyzer (ECT6008, IEST, 0.01% accuracy) and discharged with a galvanostatic discharge current of 2 mA and a cutoff voltage of 0.86 V, 0.5 V and 0.01 V. After discharging process, the cell was disassembled and the $Ti_3C_2Cl_2$ or $Ti_3C_2Br_2$ work electrode was taken out for further washing by acetone, anhydrous ethanol and deionized water to remove residual electrolyte and byproducts. Then the fresh sediment was further exfoliated by ultrasonication in ice bath. After centrifugation at 5000 rpm for 30 min, a stable few-layer $Ti_3C_2Cl_2$ or $Ti_3C_2Br_2$ suspension could be obtained. And the membrane was prepared by vacuum filtration on Celgard 3501 membrane (0.25 μm, 50 mm in diameter), and then dried at 60 °C under vacuum for 8 h.

**Synthesis of few-layer MXenes with mixed $T_x$ (F, OH, O) termination**

2g of $Ti_3AlC_2$ powder was added into 100 mL of 10 wt% HF solution, followed by reaction at 25 °C for 24 h. After washing with DI water until pH up to ~6, collected sediments were mixed with TBAOH solution for intercalation and further exfoliated by ultrasonication. After centrifugation at 5000 rpm for 30 min, a few-layered $Ti_3C_2T_x$ suspension could be obtained.

**Preparation and electrical measurement of tribovoltaic nanogenerators**

The structure of the designed semiconductor-semiconductor tribovoltaic nanogenerator (SS-TVNG) consists of p-type and n-type GaN epitaxial wafers as upper and lower friction surfaces. MXene droplets were added to the friction interface of the SS-TVNG, and the friction process was controlled using a linear motor. The operational parameters of the TVNG are shown in Table S1. For signal statistics, the first signal in each friction cycle was analyzed (as shown in

Figure S16), and all were positive. Table S2 shows the concentrations of different MXene dispersions and their dilutions. MXene dispersions were prepared in anhydrous ethanol and deionized water (DI) and tested at related ratios of 20%, 40%, 60%, 80%, and 100%. High concentration -F group dispersions were diluted and tested.

**Characterizations**

The morphologies of materials were characterized by a thermal field emission scanning electron microscope (SEM, Thermo Scientific, Verios G4 UC) equipped with an energy dispersive spectroscopy (EDS) system. X-ray diffraction (XRD) analysis of the products was performed by using a Bruker D8 ADVANCE X-ray diffractometer with Cu Kα radiation and in-situ XRD analysis by Haoyuan, DX-2700BH, China with Cu Kα radiation ($\lambda = 1.5406$ Å). The morphology and crystalline lattice were investigated by high resolution transmission electron microscope (HR-TEM, Talos F200x). The height distributions were measured by Atomic Force Microscope (AFM, Dimension Icon, Bruker). In-situ differential electrochemical mass spectrometry (DEMS, IEST) coupled with galvanostatic charge-discharge analyzer (ECT6008, IEST, 0.01% accuracy) was obtained to analyze the products of decomposed electrolyte in the electrolytic cell. Fourier transform infrared spectroscopy measurements (FTIR, Bruker INVENIOR) were utilized to analyze components. Dynamic light scattering (DLS) and zeta potential (ζ) were measured on a Zetasizer Nanoseries (Malvern Instruments Ltd). Ultraviolet photoelectron spectroscopy (UPS) characterization was conducted on Shimadzu AXIS SUPRA+. Ultraviolet-visible absorption spectroscopy was recorded on a Lambda 1050+ spectrophotometer. The water contact angles on various few-layer membranes were detected by a contact angle apparatus (DCAT21, Ningbo Jin Mao Import & Export Co., Ltd.). The electrical conductivity was measured using a four-point probe set-up (Model 280 DI, Four dimensions, Inc). Electrical characterization of the TVNG output was acquired via a Tektronix Keithley 6517B Electrostatic Meter. I-V curves were recorded on a Tektronix Keithley 4200-SCS Semiconductor Characterization System. Optical characterization of the TVNG surface was performed with a ZEISS Axio Imager m2M optical microscope.

## Additional information

**Supplementary information** The online version contains supplementary material available at

https://doi.org/

## Acknowledgments

This work was financially supported National Natural Science Foundation of China (U23A2093, and 12375279), High-Level Talents Special Support Program of Zhejiang Province (No. 2022R51007), Ningbo Top-talent Team Program, and Youth Science and Technology Innovation Leading Talent Project of Ningbo. K.L. gratefully acknowledges financial support from Anglo American Resources Trading (China) Co., Ltd.

## Competing interests

K.L., M.C. and Q.H. are inventors on patent application (file No. 2024060600832390) submitted by Ningbo Institute of Materials Technology and Engineering that cover the electrochemical exfoliation method for fabricating few-layer MS-MXenes with high efficiency described in this paper.

## Data Availability

The data that support the findings of this study are available from the corresponding author upon reasonable request.

## Keywords

MXenes, Molten salt etching, Intercalation, Electrochemical exfoliation, Tribovoltaic nanogenerators

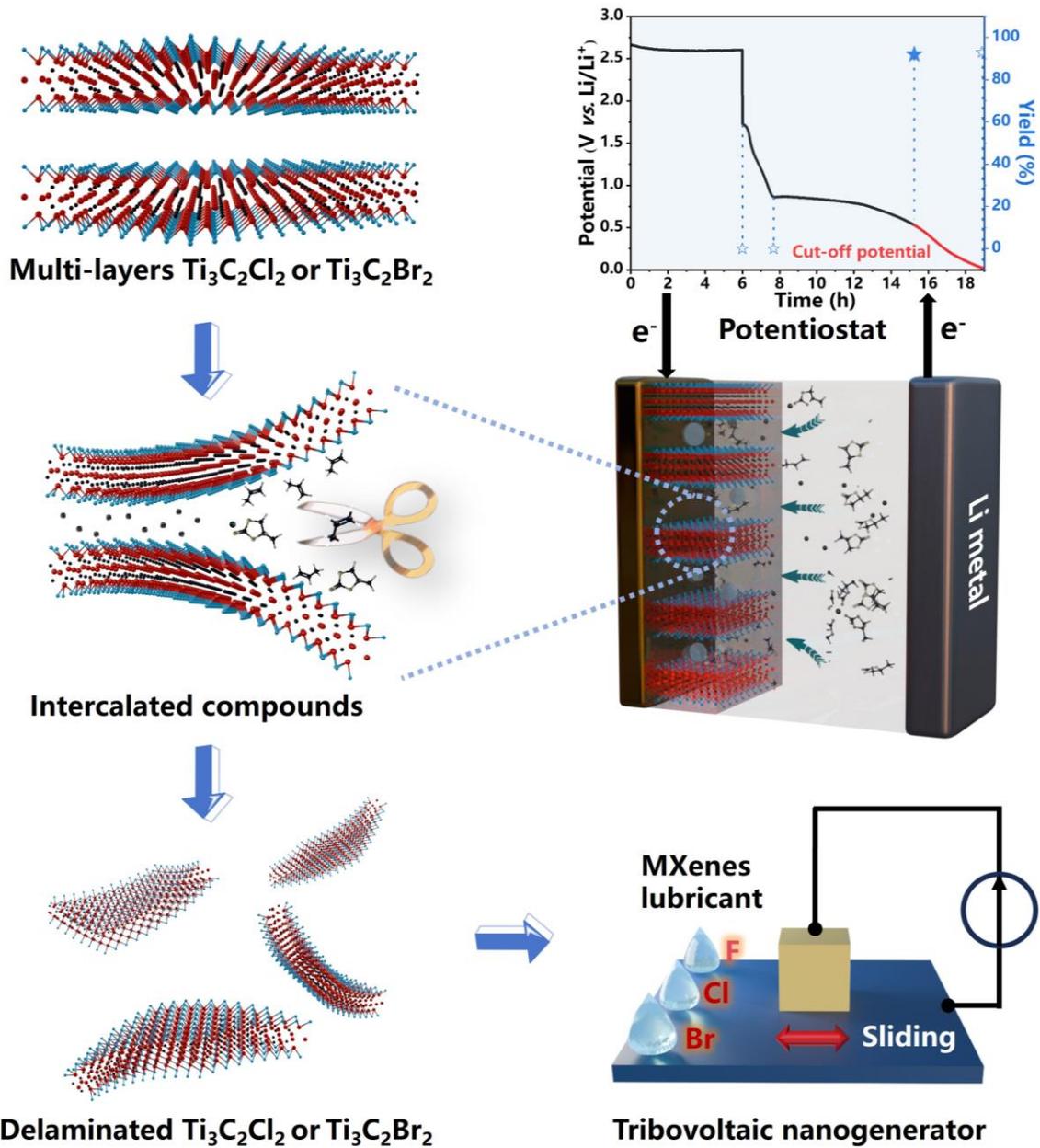

Fig. 1 Schematic illustrations of the gaseous scissor-mediated electrochemical exfoliation process for $Ti_3C_2Cl_2$ or $Ti_3C_2Br_2$ MXene nanosheets, with their applications in tribovoltaic nanogenerator.

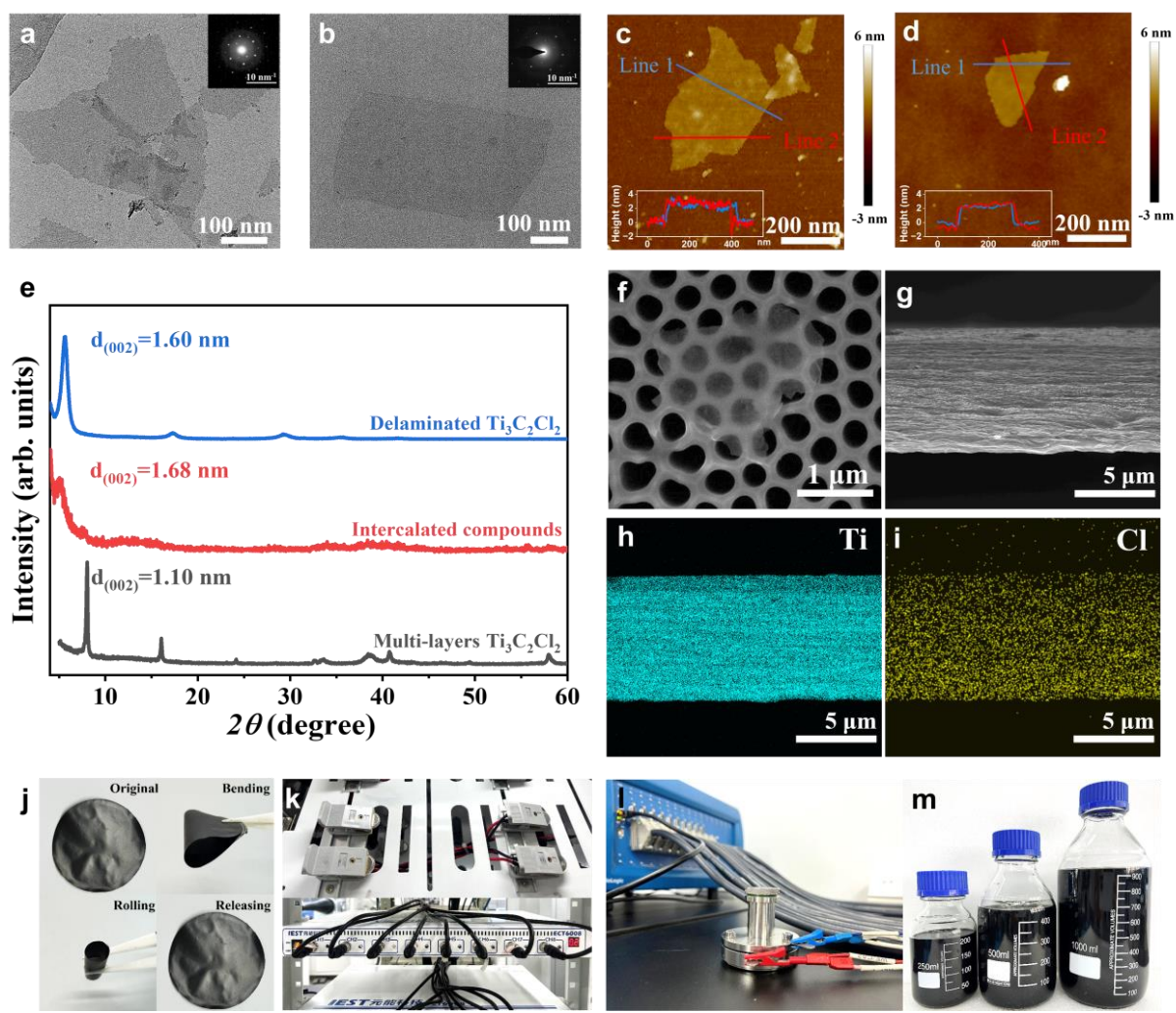

**Fig. 2 Characterizations of delaminated MS-MXenes.** TEM images of few-layer **a** $Ti_3C_2Cl_2$ and **b** $Ti_3C_2Br_2$ nanosheets, inset: corresponding selected area electron diffraction (SAED) patterns. AFM images of few-layer **c** $Ti_3C_2Cl_2$ and **d** $Ti_3C_2Br_2$ nanosheets. **e** XRD patterns of multi-layers $Ti_3C_2Cl_2$, intercalated compounds and delaminated $Ti_3C_2Cl_2$. **f** SEM image of few-layer $Ti_3C_2Cl_2$ nanosheets on an AAO substrate. Cross-sectional **g** SEM image and corresponding mapping images of **h** Ti and **i** Cl for membrane made by few-layer $Ti_3C_2Cl_2$ nanosheets. **j** Digital photograph of a piece of delaminated $Ti_3C_2Cl_2$ paper. The electrochemical lithium ion-intercalation experimental setup: **k** IEST battery testing system and **l** potentiostat. **m** Optical image of delaminated MS-MXenes solutions.

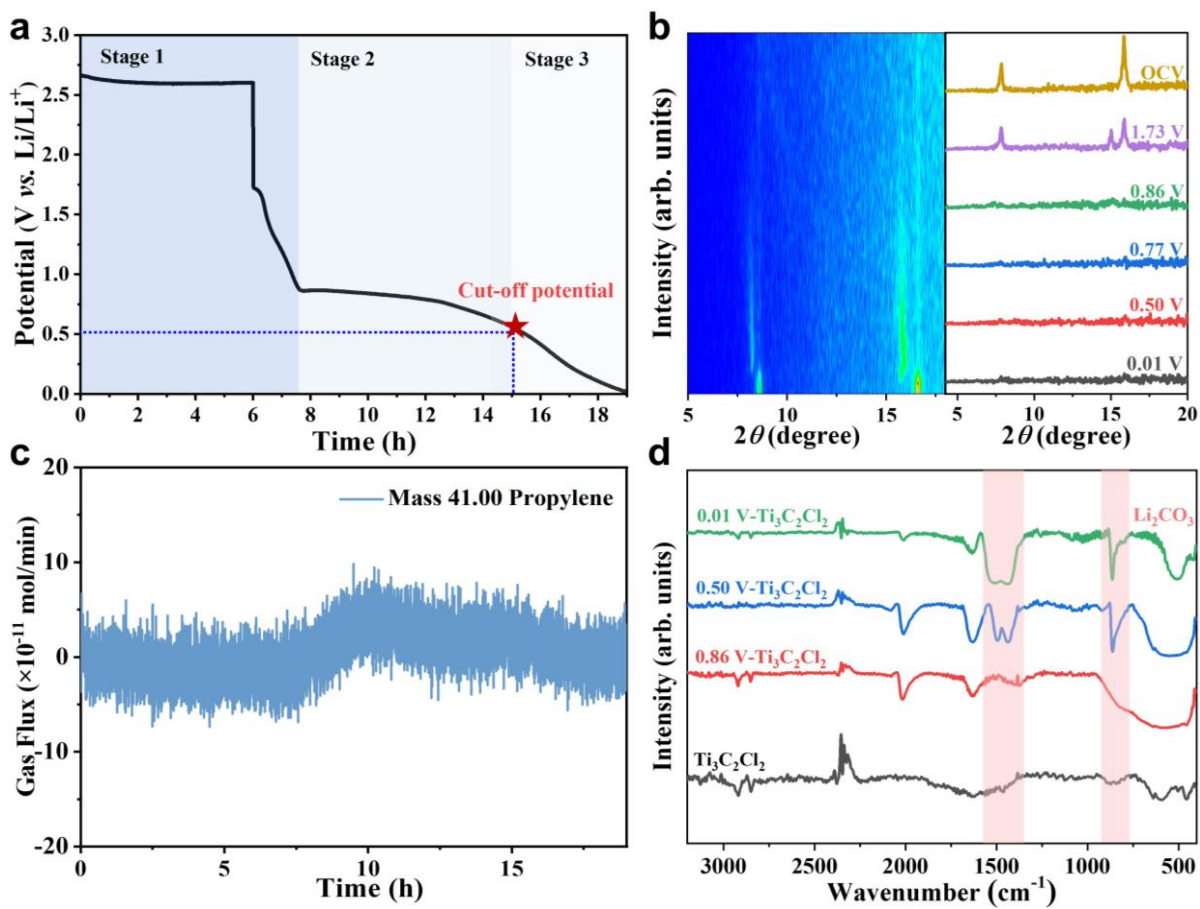

**Fig. 3 Characterizations of electrochemical exfoliation process. a** Discharge curve of the fabricated half-cell with multi-layers $Ti_3C_2Cl_2$ pellet as work electrode, Li metal as counter electrode and 1 M LiTFSI in PC as electrolyte. **b** In-situ XRD patterns (left panel) for $Ti_3C_2Cl_2$ upon electrochemical intercalation process and XRD curves at specific cut-off potentials (right panel). **c** In-situ DEMS analysis of gaseous products in the half-cell. **d** Ex-situ FT-IR spectra of the pristine $Ti_3C_2Cl_2$ and intercalated compounds at various cut-off potentials.

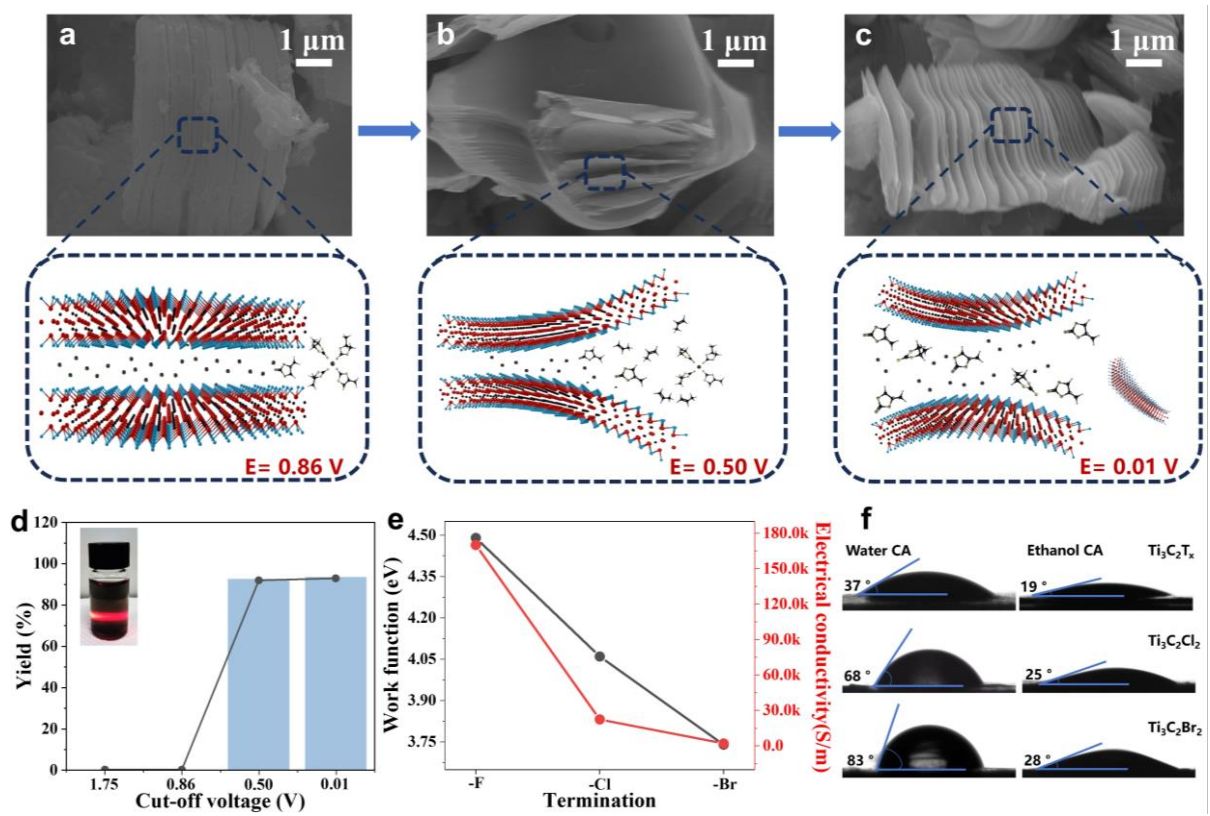

**Fig. 4 Mechanisms of electrochemical exfoliation process.** SEM images at specific discharge cut-off potentials: **a** 0.86 V, **b** 0.50 V and **c** 0.01 V, with insets showing schematic diagrams of the intercalation reaction mechanism. **d** Yield (%) of few-layered $Ti_3C_2Cl_2$ nanosheets at different cut-off potentials, with inset displaying the Tyndall effect in the delaminated $Ti_3C_2Cl_2$ solution. **e** Work function and electrical conductivities of MXenes with different terminations. **f** Water contact angles (CA) and ethanol contact angles on the HF-etched $Ti_3C_2T_x$ freestanding film (top), $Ti_3C_2Cl_2$ freestanding film (middle), and $Ti_3C_2Br_2$ freestanding film (bottom).

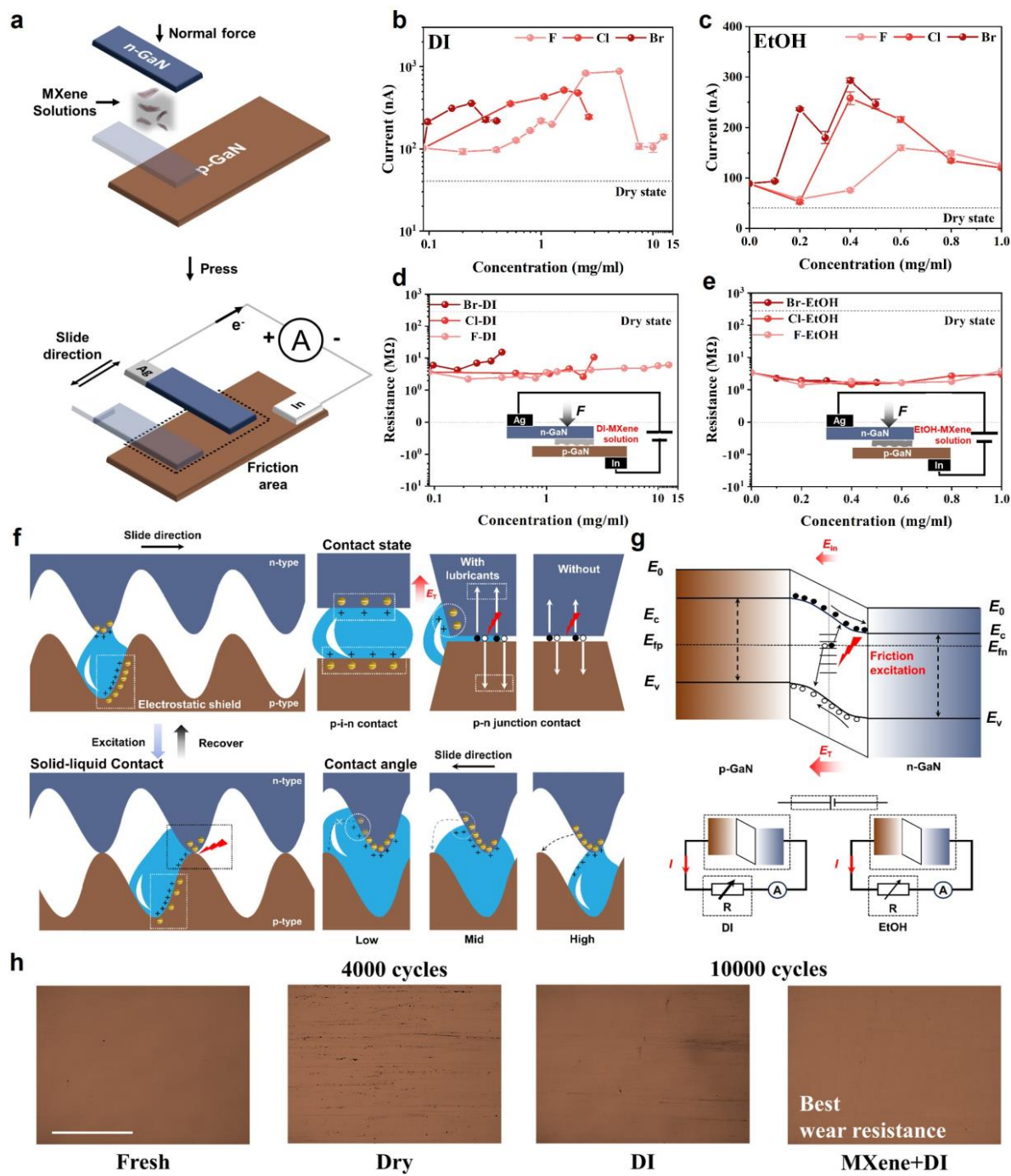

**Fig. 5 Effects of different MXenes lubricants on output performance of TVNGs. a** 3D structure of TVNG and its external circuit connection diagram. **b, c** Short-circuit current under varying concentrations of MXene with different terminations in b) water or c) ethanol solutions. **d, e** Static resistance of TVNG under a bias of 4 V for MXenes with different terminations in **d** water or **e** ethanol solutions. **f** Mechanism diagram illustrating the enhancement effect on SS-TVNG induced by solid-liquid contact. **g** Band structure diagram of the p-n junction in the device, showing the influence of frictional electric field and interface resistance. **h** Optical images of fresh TVNGs, and TVNGs after stability tests without and with MXene lubricants (Scale bar: 500 μm).

# Supplementary Information

**Gaseous Scissor-mediated Electrochemical Exfoliation of Halogenated MXenes and its Boosting in Wear-Resisting Tribovoltaic Devices**

Qi Fan et al.

**Supplementary Figures**

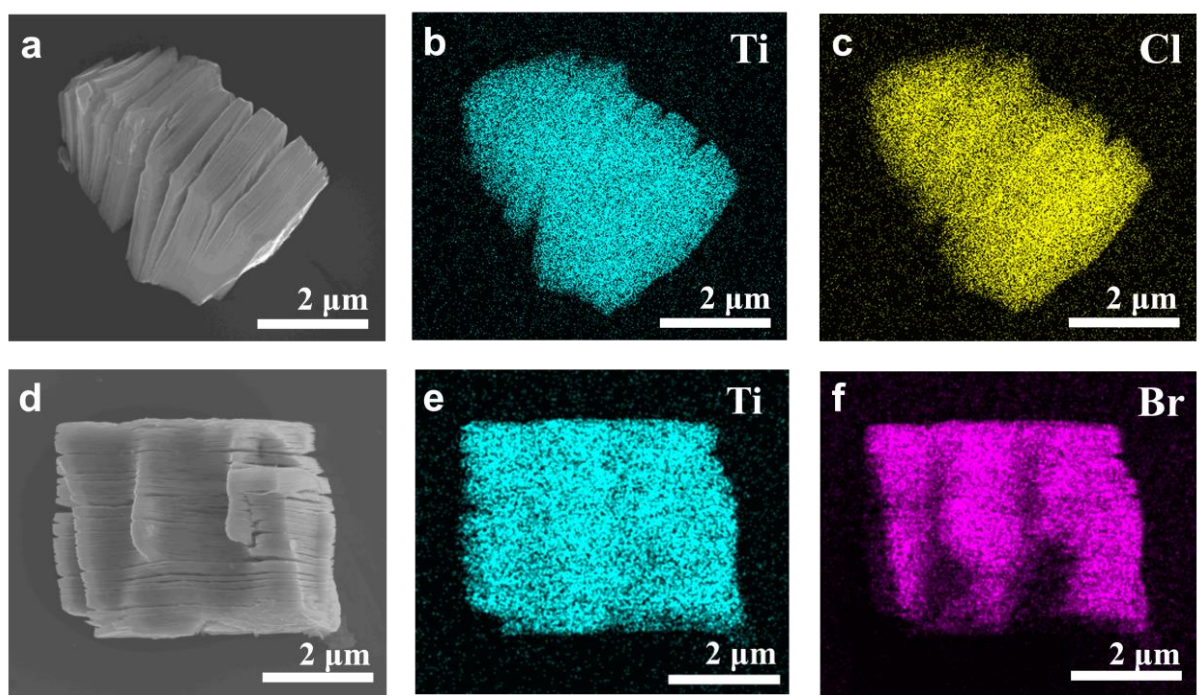

**Supplementary Figure 1.** SEM images of **a** Ti$_3$C$_2$Cl$_2$ and **c** Ti$_3$C$_2$Br$_2$; corresponding EDS mappings with the element distributions of **b** Ti, **c** Cl for Ti$_3$C$_2$Cl$_2$, and **e** Ti, **f** Br for Ti$_3$C$_2$Br$_2$.

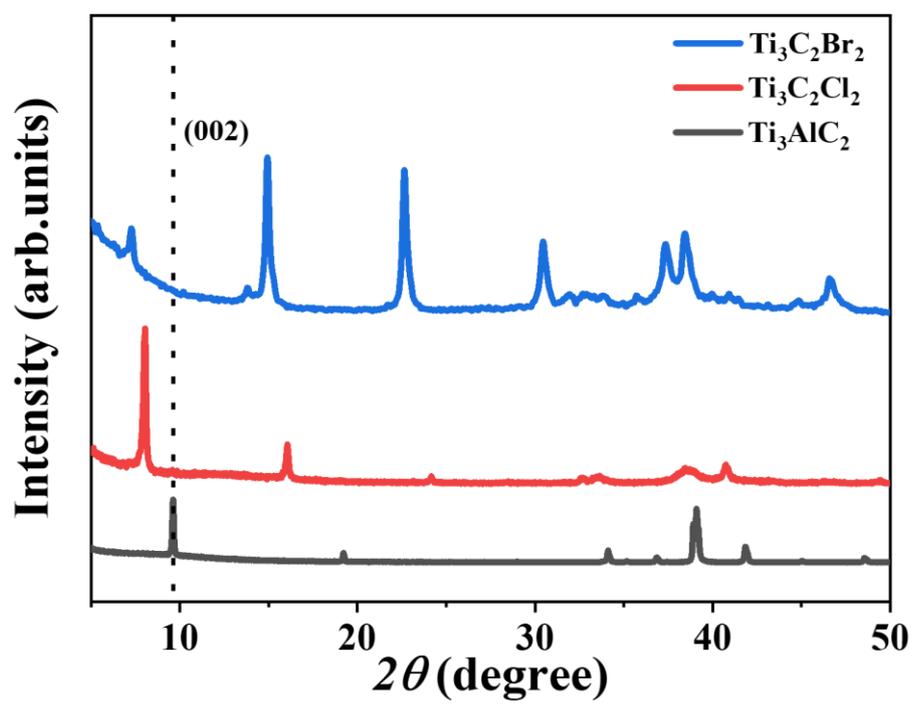

**Supplementary Figure 2.** XRD patterns of Ti$_3$AlC$_2$, the prepared Ti$_3$C$_2$Cl$_2$ and Ti$_3$C$_2$Br$_2$.

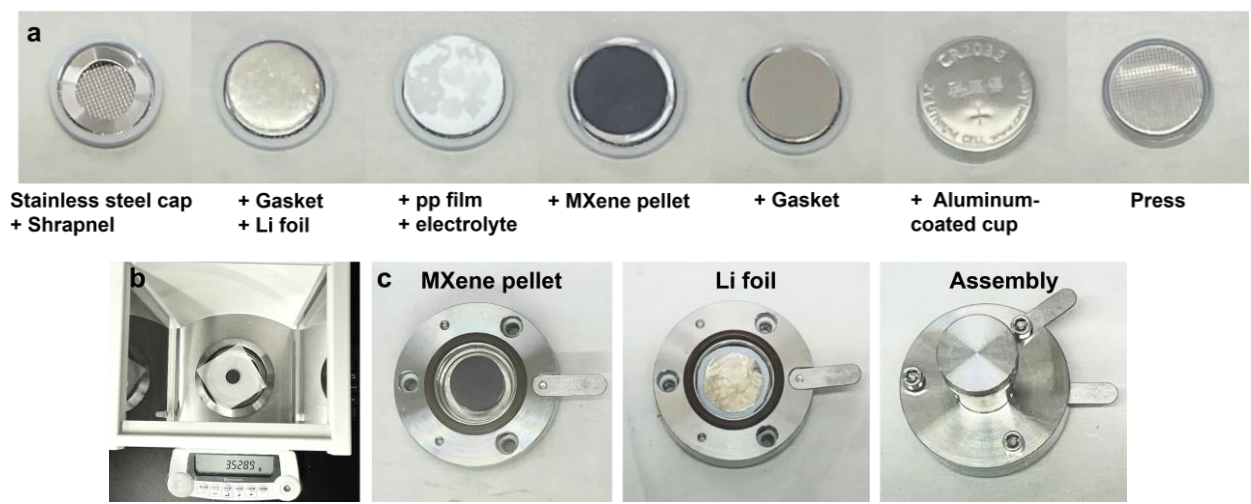

**Supplementary Figure 3.** Photographs of **a** coin cell and **c** two-electrode TMAX-2E cell assembly procedure for electrochemical intercalation. **b** Optical image of a MXene pellet with mass of ~3.5 g.

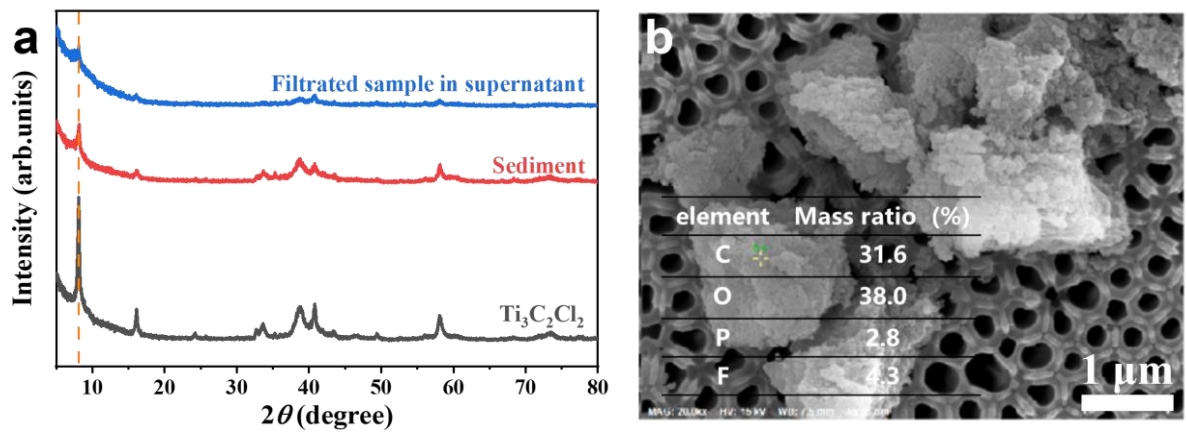

**Supplementary Figure 4. a** XRD patterns for the pristine Ti$_3$C$_2$Cl$_2$, the bottom solid and products from top solution after electrochemical intercalation followed by cleanout and filtration process. **b** SEM image and corresponding EDS results of MS-MXenes after lithiation process.

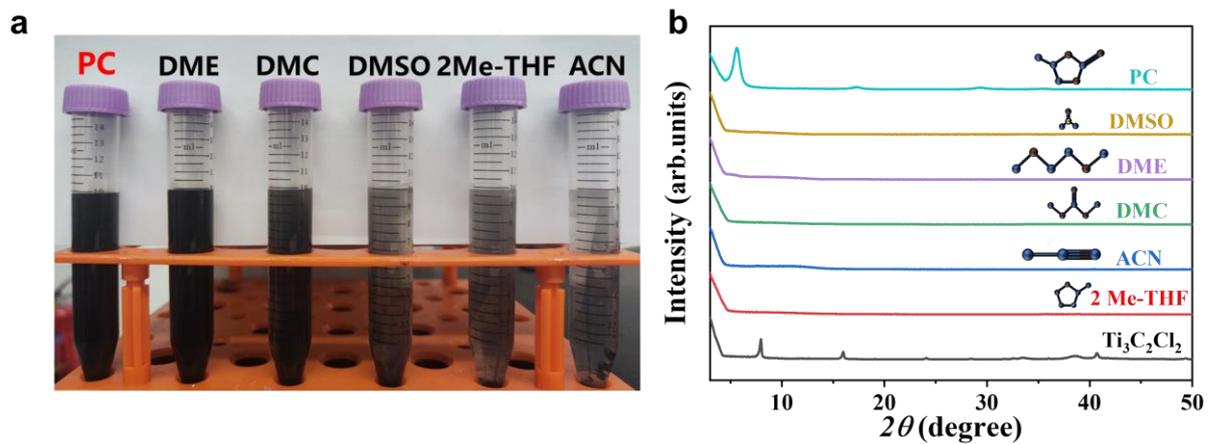

**Supplementary Figure 5. a** Digital photographs of the collected delaminated MS-MXene solutions after 4 weeks. **b** XRD patterns of exfoliated products derived from electrochemical intercalation assisted exfoliation by using various solvents.

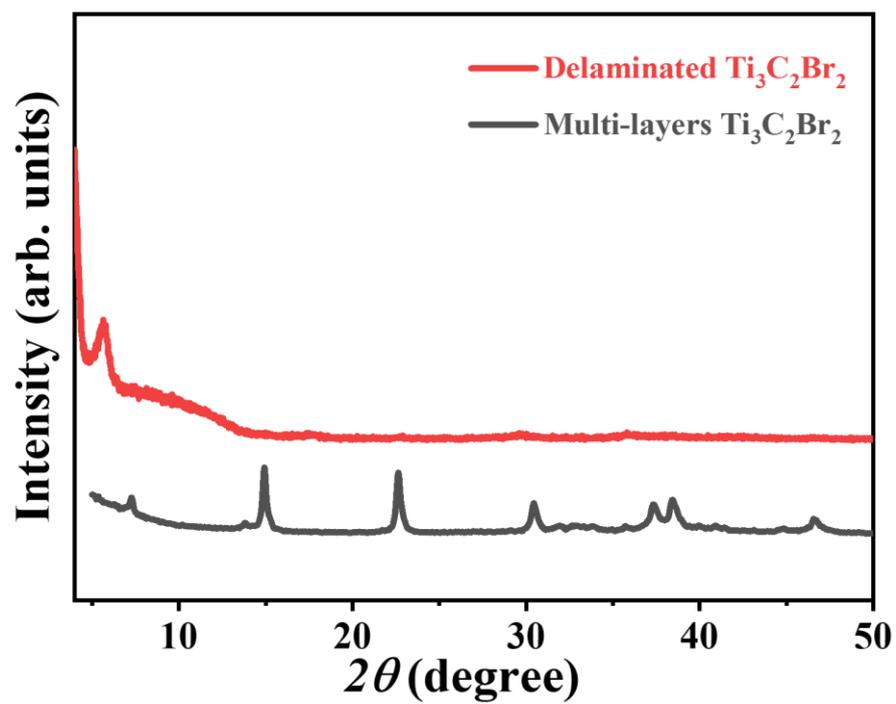

**Supplementary Figure 6.** XRD patterns of multi-layers $Ti_3C_2Br_2$ and delaminated $Ti_3C_2Br_2$.

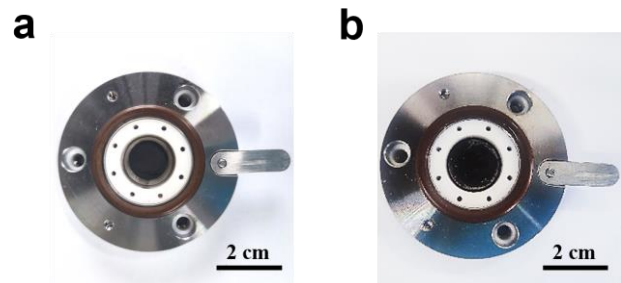

**Supplementary Figure 7.** Digital photographs of **a** multi-layers Ti$_3$C$_2$Cl$_2$ pellet and **b** intercalated compounds after electrochemical intercalation process.

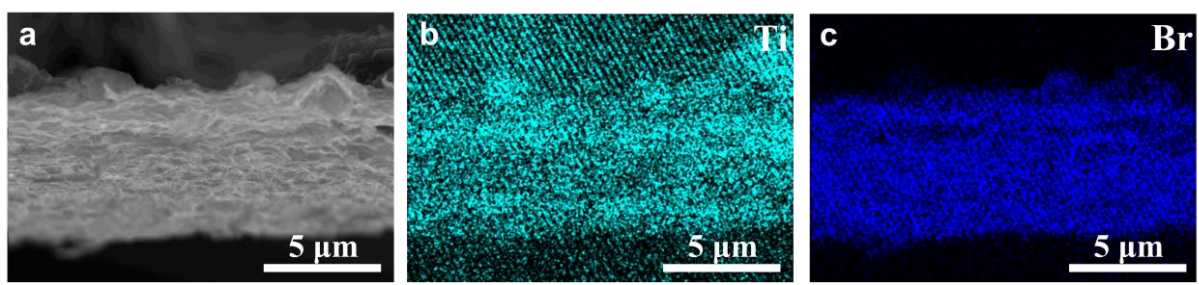

**Supplementary Figure 8.** The cross-sectional **a** SEM image and corresponding mapping images of **b** Ti and **c** Br for few-layer $Ti_3C_2Br_2$ membrane.

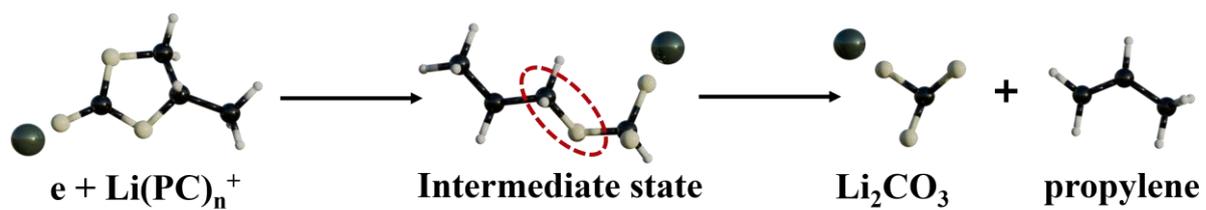

**Supplementary Figure 9.** Optimized simulation of possible reductive decomposition process of Li(PC)$_n^+$ under potentials of 0.50 V-0.86 V.

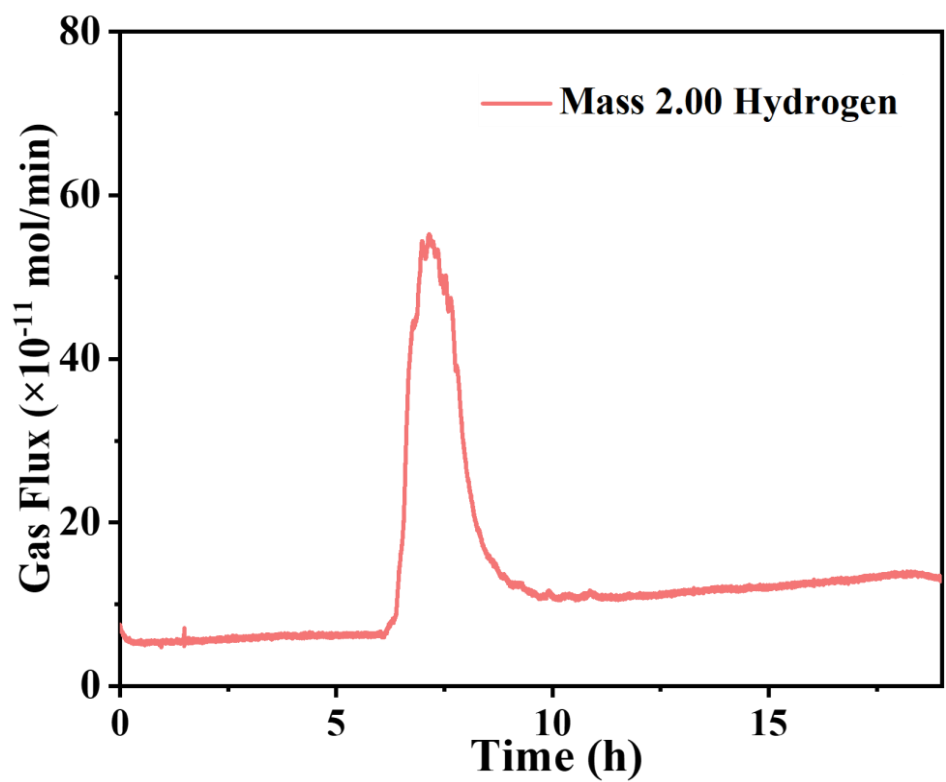

**Supplementary Figure 10.** In-situ DEMS analysis of gaseous products in the half-cell.

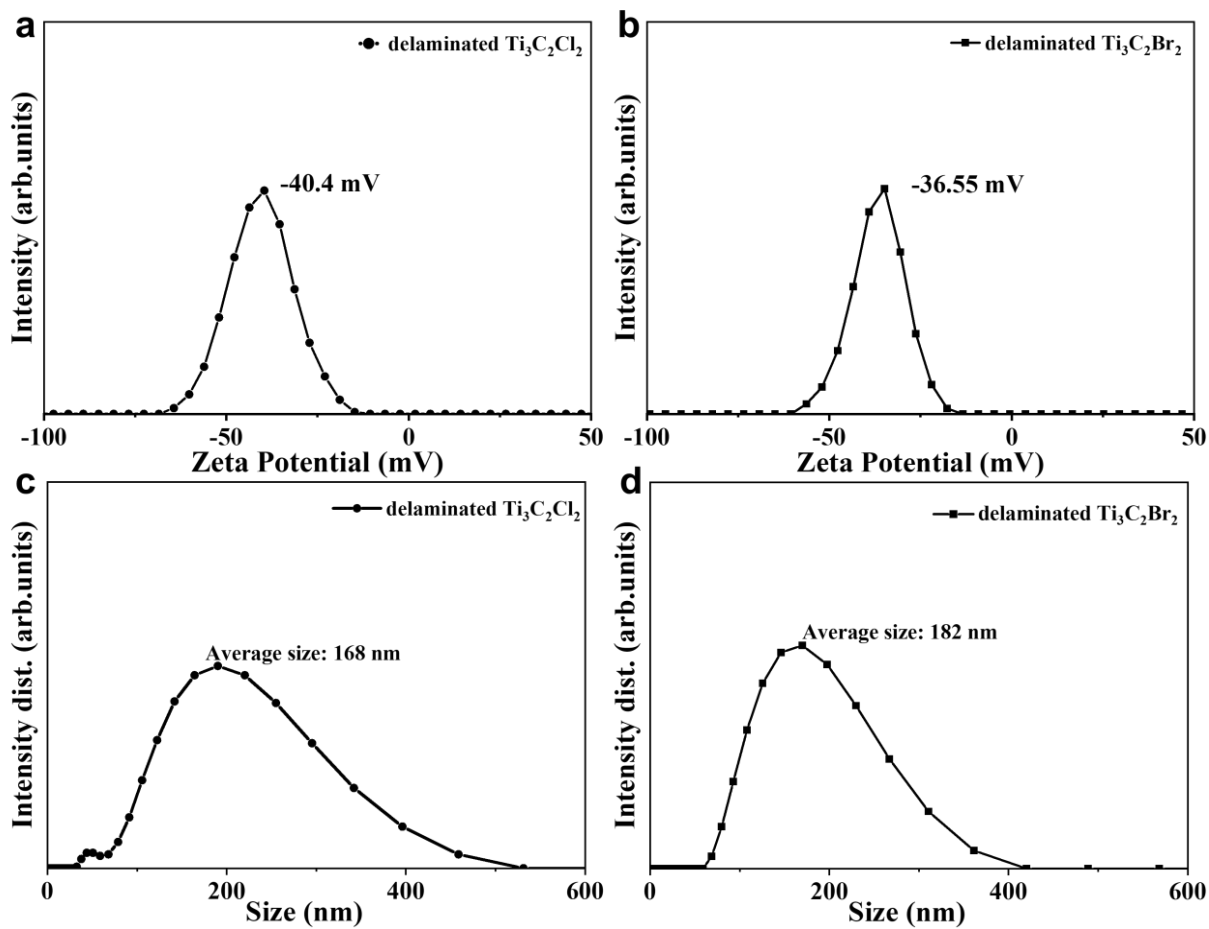

**Supplementary Figure 11.** Zeta potential of **a** delaminated $Ti_3C_2Cl_2$ and **b** $Ti_3C_2Br_2$ in aqueous solution. Mean size of **c** delaminated $Ti_3C_2Cl_2$ and **d** $Ti_3C_2Br_2$.

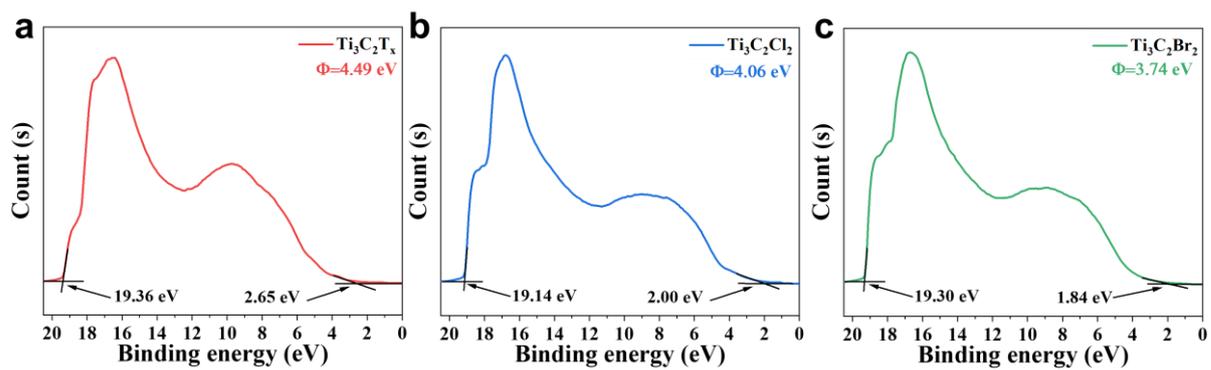

**Supplementary Figure 12.** UPS spectra of **a** Ti$_3$C$_2$T$_x$, **b** Ti$_3$C$_2$Cl$_2$, and **c** Ti$_3$C$_2$Br$_2$.

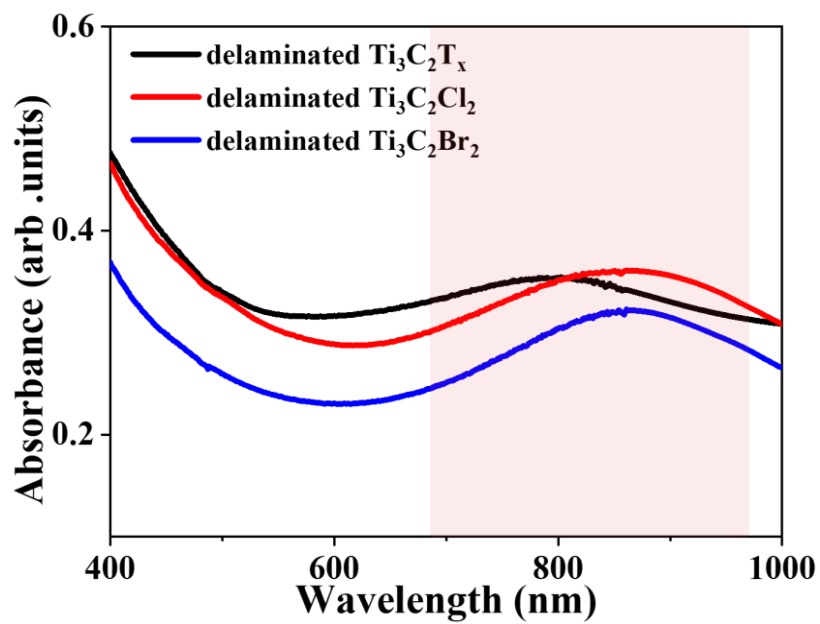

**Supplementary Figure 13.** UV-vis spectra of delaminated $Ti_3C_2T_x$, $Ti_3C_2Cl_2$, and $Ti_3C_2Br_2$ solutions.

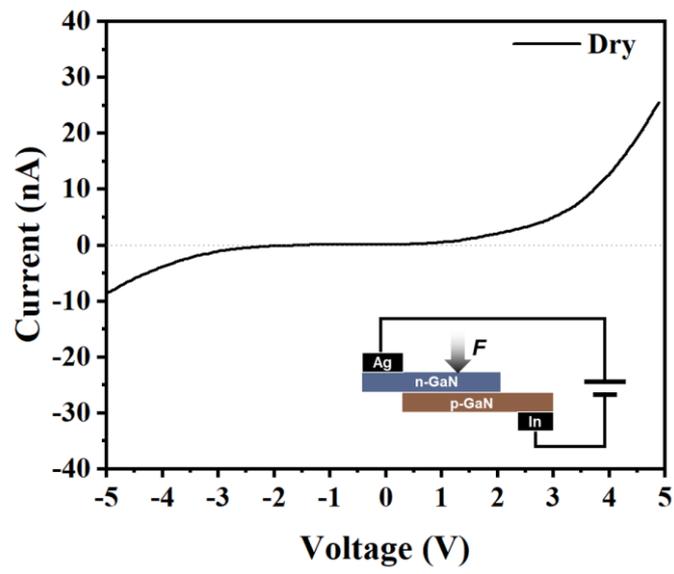

**Supplementary Figure 14.** I-V curve of TVNG without lubricant.

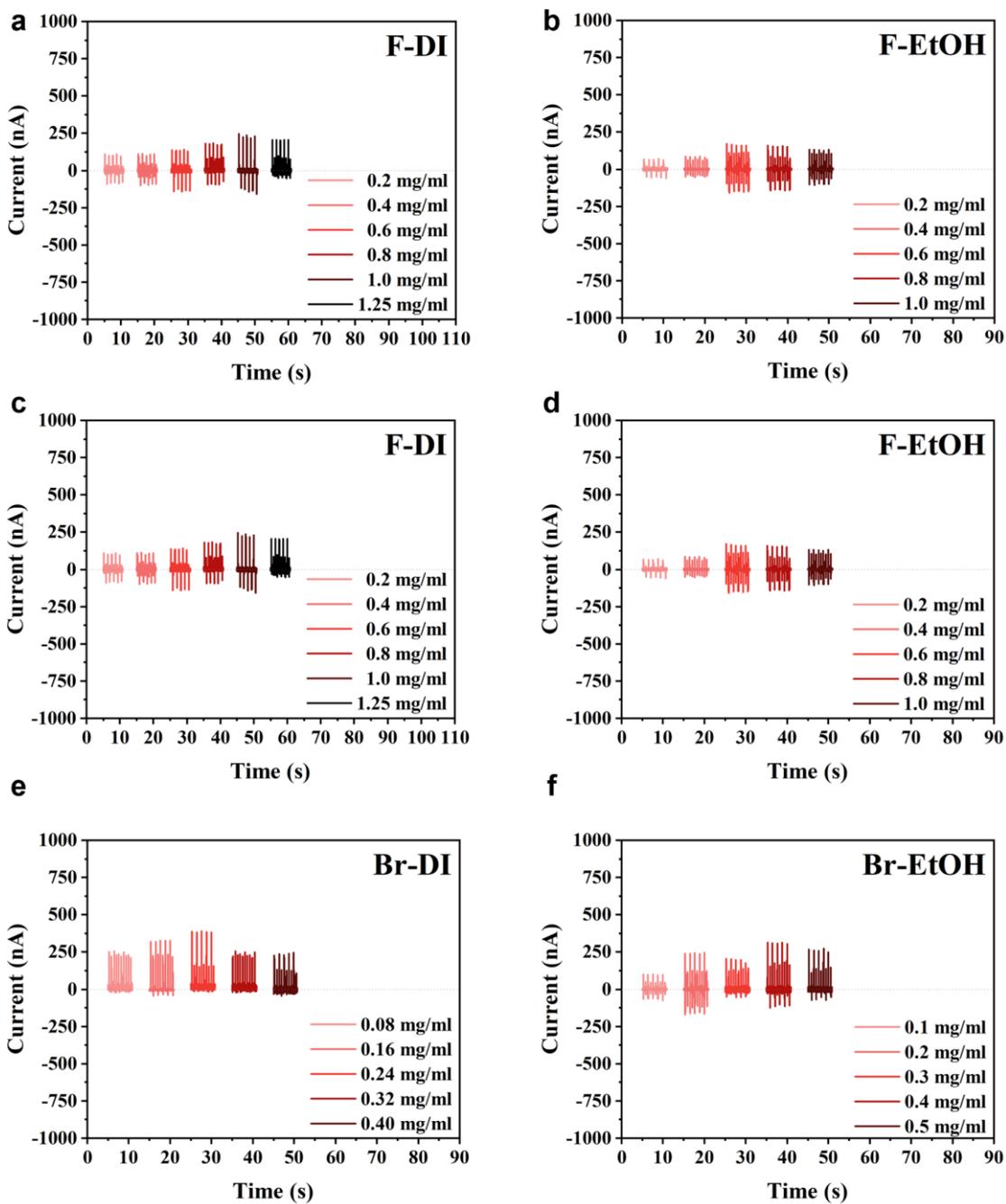

**Supplementary Figure 15.** Current output of TVNG lubricated by MXenes with different terminations in deionized water or ethanol solutions.

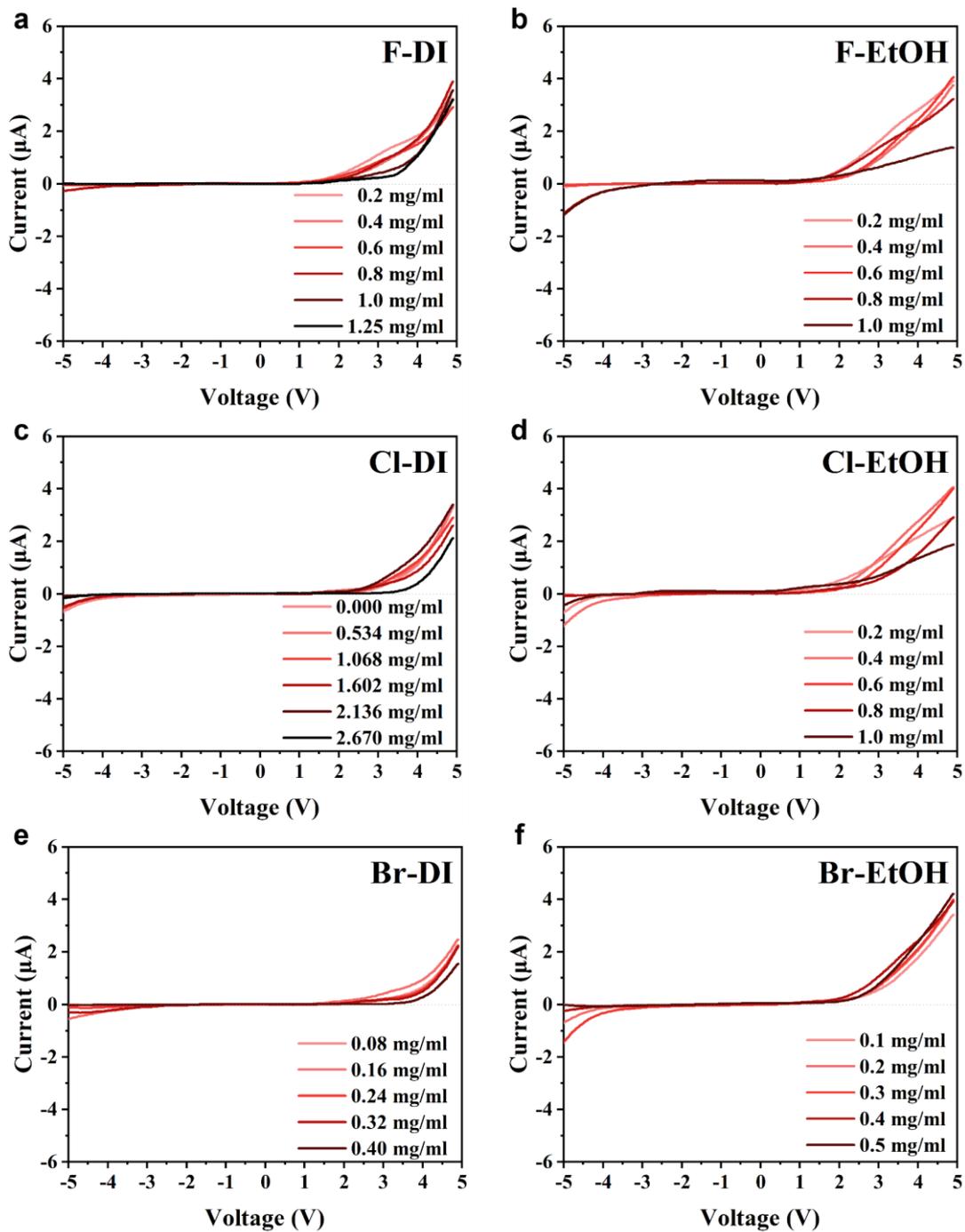

**Supplementary Figure 16.** I-V characteristic of the TVNG lubricated by MXenes with different terminations in deionized water or ethanol solutions.

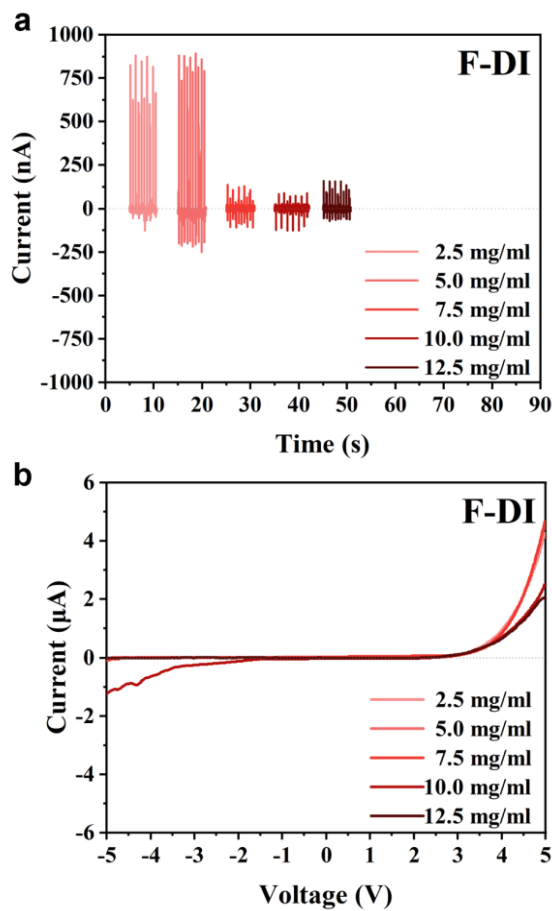

**Supplementary Figure 17. a** Current output and **b** I-V characteristic of TVNG lubricated by -F terminated MXene.

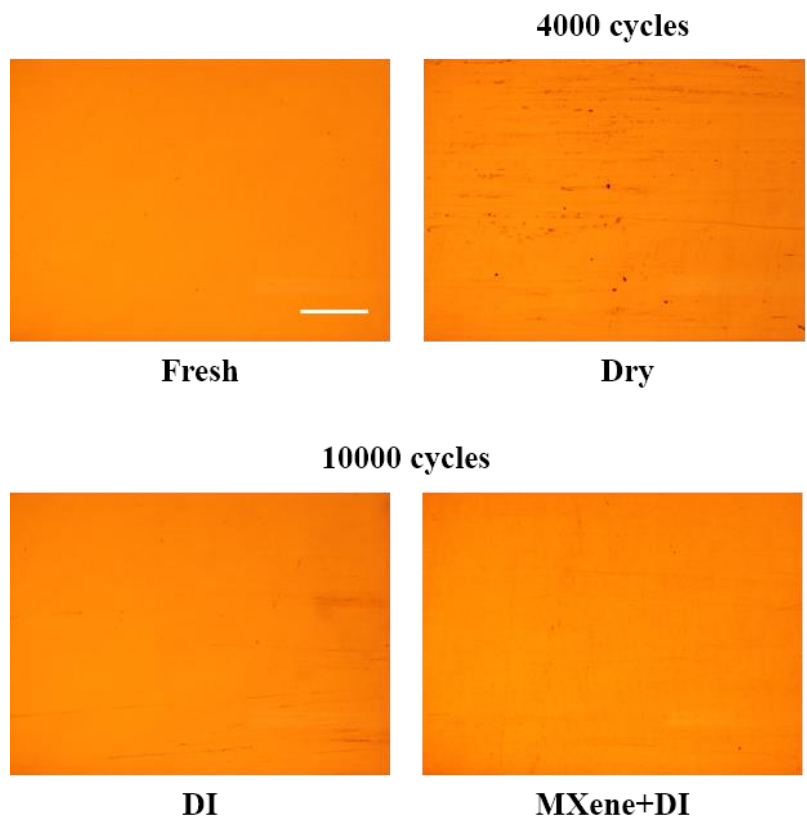

**Supplementary Figure 18.** Optical images of fresh TVNGs, and TVNG after stability tests without and with MXene lubricants, at lower magnification (Scale bar: 500 μm).

**Supplementary Table 1.** The operating parameters of SS-TVNG.

| | |
|---|---|
| Acceleration | 1 m/s² |
| Deceleration | 1 m/s² |
| Velocity | 0.1 m/s |
| Start | -19.065 mm |
| End | -26.558 mm |
| Distance | 7.493 mm |
| Normal Force | 10 N |
| Interval Time | 500 ms |

**Supplementary Table 2.** MXene dispersions used in experiments.

| Dispersion | Medium | Initial Concentration mg/ml | 20% mg/ml | 40% mg/ml | 60% mg/ml | 80% mg/ml |
|---|---|---|---|---|---|---|
| EtOH | F | 1.0 | 0.2 | 0.4 | 0.6 | 0.8 |
| | Cl | 1.0 | 0.2 | 0.4 | 0.6 | 0.8 |
| | Br | 0.5 | 0.1 | 0.2 | 0.3 | 0.4 |
| DI | F | 12.5 | 2.5 | 5.0 | 7.5 | 10.0 |
| | | 1.0 | 0.2 | 0.4 | 0.6 | 0.8 |
| | Cl | 2.67 | 0.534 | 1.068 | 1.602 | 2.136 |
| | Br | 0.4 | 0.08 | 0.16 | 0.24 | 0.32 |